\documentclass[prb,twocolumn,showpacs,floatfix,superscriptaddress]{revtex4}
\usepackage{graphicx}
\usepackage{dcolumn}
\usepackage{bm}
\usepackage{color}

\begin{document}

\title{Partially-ordered vortex lattices in the high-field, low-temperature
mixed state of quasi two-dimensional organic superconductors }
\author{A. Maniv}
\affiliation{NRCN, P.O. Box 9001, Beer Sheva, 84190, Israel}
\author{V. Zhuravlev}
\affiliation{Schulich Faculty of Chemistry, Technion-Israel Institute of Technology,
Haifa 32000, Israel}
\author{T. Maniv}
\email{maniv@tx.technion.ac.il}
\affiliation{Schulich Faculty of Chemistry, Technion-Israel Institute of Technology,
Haifa 32000, Israel}
\author{O. Ofer}
\affiliation{Schulich Faculty of Chemistry, Technion-Israel Institute of Technology,
Haifa 32000, Israel}
\author{R. Rommel}
\affiliation{Institute of Physics, SFB/TR49, Goethe University Frankfurt,
Max-von-Laue-Str. 1, 60438 Frankfurt am Main, Germany}
\author{J. M\"{u}ller}
\affiliation{Institute of Physics, SFB/TR49, Goethe University Frankfurt,
Max-von-Laue-Str. 1, 60438 Frankfurt am Main, Germany}
\author{J. E. Sonier}
\affiliation{Department of Physics, Simon Fraser University, Burnaby, British Colombia
V6T 1Z1, Canada}
\affiliation{Canadian Institute for Advanced Research, Toronto, Ontario M5G 1Z8, Canada}
\date{\today }

\begin{abstract}
We report the results of high-field, low-temperature $\mu $SR measurements
of the quasi two-dimensional organic superconductors $\kappa $-(ET)$_{2}$%
Cu(NCS)$_{2}$ and $\kappa $-(ET)$_{2}$Cu[N(CN)$_{2}$]Br. The $\mu $SR
lineshapes for these compounds indicate the existence of partially-ordered
vortex lattice phases in the high magnetic field regime, up to 2.5 T for the
former compound and 4 T for the latter compound. The observed sharp loss of
order is found to be consistent with a vortex-lattice melting transition
that is predicted by numerical simulations of weakly coupled layers of
pancake vortices. It is argued that the robustness of the partially-ordered
vortex lattice phases could be due to strong flux-line pinning by a dilute
ensemble of defects.
\end{abstract}

\pacs{74.70.Kn, 74.25.Uv, 76.75.+i}
\maketitle


\section{Introduction}

In quasi two-dimensional (2D), extreme type-II superconductors, such as the
organic charge transfer salts (ET)$_{2}$X,\cite{Yamaji90,Toyota07} and the
high-transition temperature ($T_{c}$) compound Bi$_{2}$Sr$_{2}$CaCu$_{2}$O$%
_{8}$ (BSCCO),\cite{Subramanian88} the existence and extent of highly
ordered three-dimensional (3D) vortex lattice states are known to be
strongly dependent on the coupling between the underlying 2D superconducting
(SC) layers. In such compounds with weakly coupled SC layers (via very small
Josephson tunneling currents) subjected to a magnetic field applied
perpendicular to the layers,\cite{Wosnitza96} the vortex system consists of
stacked layers of pancake-like vortices.\cite{BlatterRMP94} For applied
magnetic fields not far below the upper critical field ($H_{c2}$), the
relatively strongly coupled intralayer pancake vortices can form well
ordered 2D lattices, whereas the pancake vortices in adjacent layers couple
very weakly by the electromagnetic dipole interaction. Consequently, while
at low temperatures straight magnetic flux lines threading weakly coupled
pancake vortices may essentially form a 3D ordered lattice, these flux lines
can be easily distorted by thermal fluctuations at elevated temperatures.
This leads to a loss of 3D order, which occurs at a dimensional crossover
field ($H_{\mathrm{2D}}$) far below $H_{c2}$. Further disordering of the
flux lattice, by strong thermal fluctuations within the individual layers of
vortices, can occur at much higher magnetic field, causing a complete loss
of order (vortex-lattice melting) at an intermediate field $H_{\mathrm{2D}%
}\! \ll \!H\!<\!H_{c2}$ \cite{HuMacDonald93-94,ZM99,Maniv01}. These
order-disorder transitions have been extensively studied in BSCCO, via for
example, small-angle neutron scattering (SANS),\cite{CubittNature93} muon
spin rotation ($\mu $SR),\cite{LeePRL93,LeePRB97} irreversibility-line \cite%
{SchillingPRL93} and magnetization \cite{TamegaiPhysicaC93} measurements.
Similar vortex-lattice transitions have also been investigated in the low-$%
T_{c}$, quasi-2D organic superconductors (see Ref.~\onlinecite{Toyota07} for
a review), with applications of the $\mu $SR technique focusing on the
charge transfer salts $\kappa $-(ET)$_{2}$Cu(NCS)$_{2}$,\cite%
{Lee-ET-PRL97,Blundell2004} and $\kappa $-(ET)$_{2}$Cu[N(CN)$_{2}$]Br\cite%
{Lancaster2011}.

Theoretical estimates of the 3D-2D crossover field $H_{\mathrm{2D}}$, based
on a continuum elastic description of the vortex lattice,\cite%
{GlazmanPRB91,BlatterPRB96} yield characteristic values in the range of $H_{%
\mathrm{2D}}\sim \Phi _{0}/\lambda ^{2}$ (where $\Phi _{0}$ is the magnetic
flux quantum, and $\lambda $ is the in-plane magnetic penetration depth),
which is about 3 orders of magnitude smaller than $H_{c2}$. These estimates
have been confirmed in a number of $\mu $SR experiments performed on the
relevant compounds.\cite{LeePRL93,LeePRB97,Blundell2004} However, in some
cases the loss of order deduced from changes in the $\mu $SR lineshape with
increasing magnetic field was incomplete,\cite{AgerterPRB96} indicating the
existence of partially-ordered 3D flux-line networks at magnetic fields well
above the theoretically predicted values of $H_{\mathrm{2D}}$. The existence
of such partially-ordered 3D vortex phases at magnetic fields $H\! \gg \!H_{%
\mathrm{2D}}$ has been demonstrated by transport\cite{Zeldov98} and SANS
measurements\cite{Forgan96} of BSCCO single crystals.

In the present paper we provide clear $\mu $SR evidence for the existence of
a partially-ordered vortex lattice phase in the quasi-2D organic
superconductors $\kappa $-(ET)$_{2}$Cu[N(CN)$_{2}$]Br ($\kappa $-Br) and $%
\kappa $-(ET)$_{2}$Cu(NCS)$_{2}$ ($\kappa $-CuNCS) at magnetic fields much
higher than the values of $H_{\mathrm{2D}}$ predicted by the continuum
elastic model. We also show that the observed sharp changes of the $\mu $SR
lineshape, generally associated with the loss of flux-line order, is
consistent with melting of pancake vortex lattices within the layers, rather
than a 3D-2D crossover. We suggest that in the low-temperature,
high-magnetic field region of the phase diagram investigated, small
concentrations of defects that pin flux lines perpendicular to the SC
layers, effectively resist the breakup of the entire 3D vortex lattice by
thermal fluctuations.

\section{Experimental}

Transverse-field (TF) $\mu $SR measurements with applied magnetic fields in
the range $1.0\!<\!H\!<\!4.5$~T were carried out on the M15 muon beam line
at TRIUMF, using a spectrometer consisting of a top loading Oxford
Instruments dilution-refrigerator (DR). The magnetic field was applied
perpendicular to the highly-conducting layers, i.e., parallel to the $b$%
-axis, or $a^{\ast }$-axis of each sample, which were mosaics consisting of
tens of single crystals of $\kappa $-Br or $\kappa $-CuNCS. The use of a
mosaic was necessary, due to the small size of the individual single
crystals --- typically less than $1$~mm wide and $100$~$\mu $m thick. The
crystals were mounted on a pure Ag sample holder using Apiezon grease. \ The
single crystals were grown by standard electrochemical crystallization
according to the literature methods.\cite{Urayama88,Wang90} The
superconducting transition temperature $T_{c}$ determined by thermodynamic
methods is $9.3\pm 0.1$ K and $11.5\pm 0.1$ K for $\kappa $-CuNCS and $%
\kappa $-Br, respectively.

A fast Fourier transform (FFT) of the TF-$\mu $SR signal closely resembles
the internal magnetic-field distribution $P(B)$.\cite{S13} The measurements
reported here were typically done by first cooling the sample from room
temperature to $T\!=\!20$~mK in a magnetic field of $H\!=\!1$~T, and
subsequently measuring the field dependence of the TF-$\mu $SR signal. For
each value of the applied field, measurements were also performed at $%
T\!=\!10$~K, which is well above the irreversibility lines for both
compounds,\cite{Sasaki98,Hagel97} and hence at a temperature where the
vortex structure has no influence on the $\mu $SR lineshape. The
measurements at $T\!=\!10$~K (henceforth referred to as the reference
signal), provide a visualization of the broadening of the TF-$\mu $SR
lineshape by the sample nuclear dipole moments, the field inhomogeneity of
the external magnet, and the background from muons stopping in the sample
holder and other places outside the sample.

\section{Results}

\begin{figure}[tbp]
\begin{center}
\includegraphics[width=8.5cm]{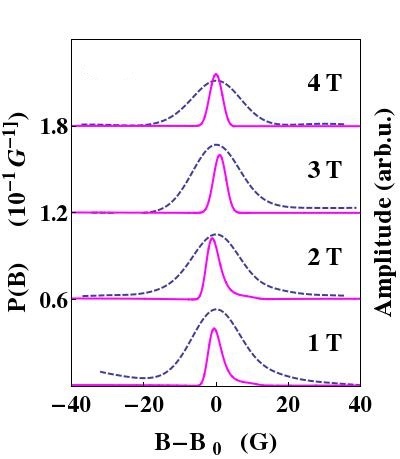}
\end{center}
\caption{(color online) Probability {magnetic field distribution}, $P(B)$,
for $\protect \kappa $-(ET)$_{2}$Cu(NCS)$_{2}$ (magenta solid {curves and}
left-hand scale) at {$T\!=\!20$~mK and} different external magnetic fields, {%
obtained} by deconvoluting the FFTs (blue dashed {curves and} right-hand
scale) of the {TF-}$\protect \mu $SR signals. {An order-disorder} transition {%
is observable between $H\!=\!2$~T and $H\!=\!3$~T.} {These measurements}
were performed after field-cooling to {$T\!=\!20$~mK at $H\!=\!1$~T.} The {%
curves are offset} vertically for {visual clarity}. The reference signal{s}
used in the deconvolution {process} (not shown) were measured on the same
sample at $T\!=\!10$~K, {well} above the irreversibility line.}
\label{Fig.1}
\end{figure}

\begin{figure}[tbp]
\begin{center}
\includegraphics[width=8.5cm]{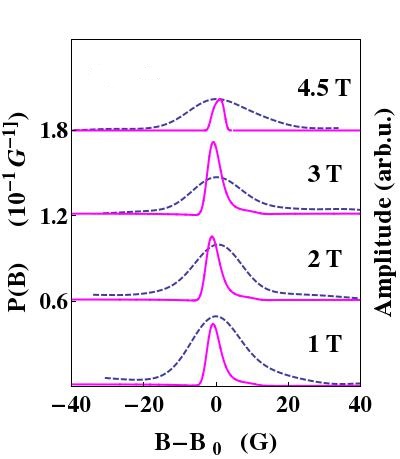}
\end{center}
\caption{(color online) Probability {magnetic field distribution}, $P(B)$,
for $\protect \kappa $-(ET)$_{2}$Cu[N(CN)$_{2}$]Br (magenta solid {curves and}
left-hand scale) at {$T\!=\!20$~mK and} different external magnetic fields, {%
obtained} by deconvoluting the FFTs (blue dashed {curves and} right-hand
scale) of the $\protect \mu $SR signals{. A} disordering transition {is
observed between $H\!=\!3$~T and $H\!=\!4.5$~T.} {These measurements} were
performed after field-cooling to {$T\!=\!20$~mK at $H\!=\!1$~T. As in Fig.~%
\protect \ref{Fig.1}, the reference signals recorded at $T\!=\!10$~K are not
shown.}}
\label{Fig.2}
\end{figure}

\begin{figure}[t]
\includegraphics[width=10cm]{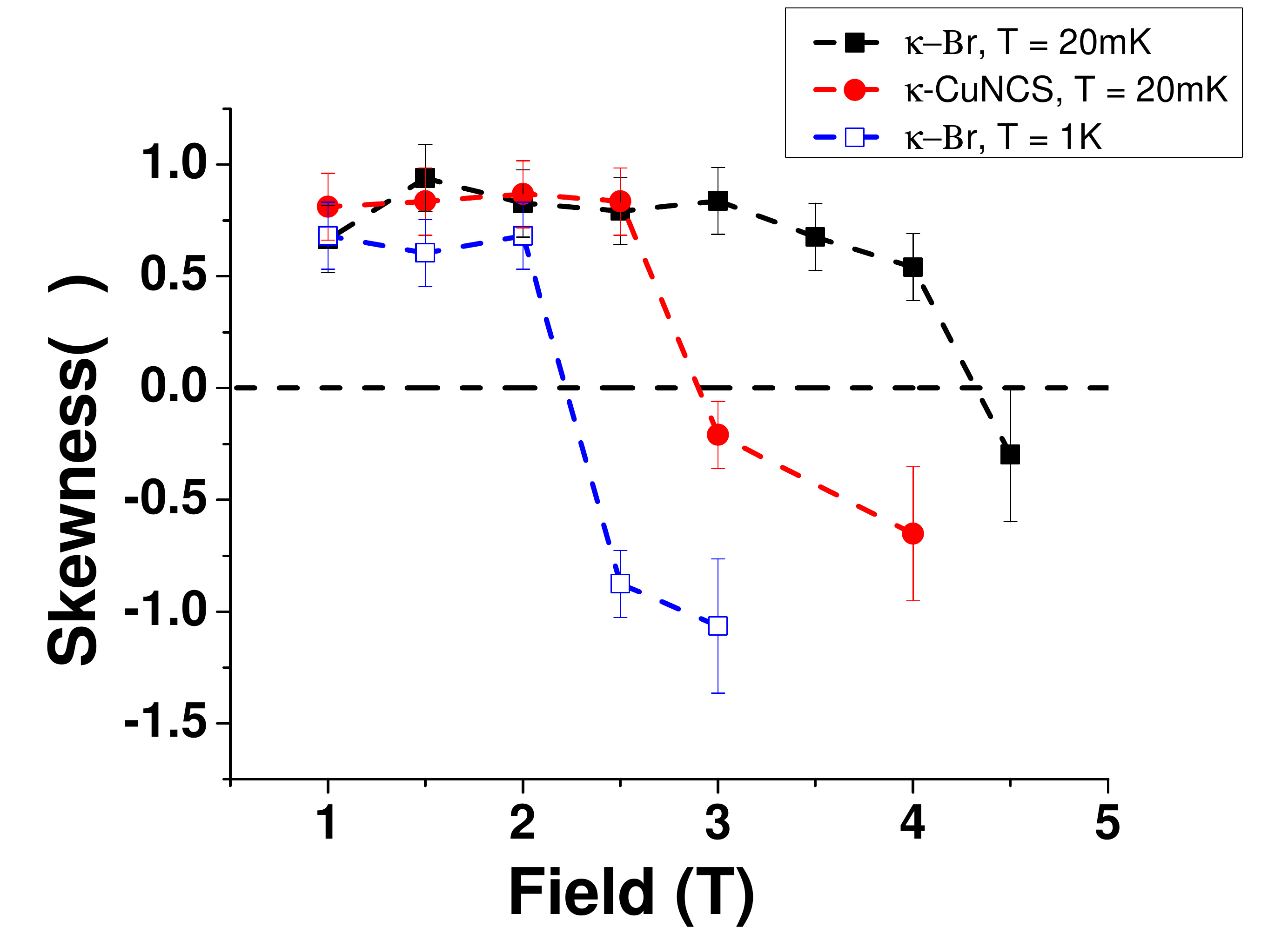}
\caption{(color online) Field dependence of the {skewness} parameter ($%
\protect \alpha $) for $\protect \kappa $-(ET)$_{2}$Cu(NCS)$_{2}$ at {$%
T\!=\!20 $~mK} (black {solid} squares), and for $\protect \kappa $-(ET)$_{2}$%
Cu[N(CN)$_{2}$]Br at {$T\!=\!20$~mK} (red {solid} circles) and {$T\!=\!1$~K}
(blue {open} squares). The lines connecting the {data} points are guides to
the eye. }
\label{Fig.3}
\end{figure}

Figure~\ref{Fig.1} shows typical FFTs of the TF-$\mu $SR signals for $\kappa 
$-CuNCS at $T\!=\!20$~mK, and the corresponding probability field
distribution, $P(B)$, obtained by deconvoluting each FFT in the frequency
domain with respect to the reference signal. Figure~\ref{Fig.2} shows
results of similar measurements for the sister compound $\kappa $-Br. \ The
most striking feature of the deconvoluted TF-$\mu $SR lineshapes shown in
Figs.~\ref{Fig.1} and \ref{Fig.2}, is their clear asymmetry, which is
characterized by a positive skewness parameter $\alpha =\left \langle \Delta
B^{3}\right \rangle ^{1/3}/\left \langle \Delta B^{2}\right \rangle ^{1/2}$,
where $\Delta B\!=\!B-\left \langle B\right \rangle $ and $\langle B\rangle $
is the average internal field\cite{S13,Maniv2011}. This is clearly observed
in a broad field range, up to a $H\!=\!2.5$~T for $\kappa $-CuNCS, and up to 
$H\!=\!4$~T for $\kappa $-Br --- above which $\alpha $ changes sign rather
abruptly. As shown in Fig.~\ref{Fig.3}, at $T\!=\!20$~mK positive values of
the skewness parameter in the range $0.5\!<\! \alpha \!<\!1.0$ reflect the
striking existence of partially-ordered 3D vortex-lattice states at magnetic
fields $H\! \gg \!H_{\mathrm{2D}}$ in both materials, which are comprised of
very weakly coupled 2D conducting layers.\cite{Singleton02} At higher
temperatures, the field range of the ordered phase shrinks considerably, as
shown in Fig.~\ref{Fig.3} for $\kappa $-Br at $T\!=\!1$~K, where the onset
of a negative $\alpha $ is observed above $H\!=\!2$~T. Note that previous TF-%
$\mu $SR lineshape measurements on $\kappa $-CuNCS performed at $T\!=\!1.8$%
~K,\cite{Lee-ET-PRL97,Blundell2004} showed positive values of the skewness
parameter $\beta =\left \langle \Delta B\right \rangle /\left \langle \Delta
B^{2}\right \rangle ^{1/2}$, which drops to zero in fields of 10 to 20~mT,
consistent with the continuum elastic theory.\cite{GlazmanPRB91,BlatterPRB96}
\ As we discuss in the next section, the rapid pre-cooling of the samples
that occurred when they were loaded into the dilution refrigerator may have
resulted in intrinsic structural disorder. Pinning associated with this
disorder may be responsible for the observed robustness of the vortex
lattice to thermal fluctuations up to high magnetic fields.

\section{Discussion}

For the first time, the high-field, low-temperature region of the vortex
state of the quasi-2D organic superconductors, $\kappa $-CuNCS and $\kappa $%
-Br, have been investigated by $\mu $SR. At the lowest temperature
considered ($T\!=\!20$~mK), an order-disorder crossover field, as determined
from the sign change of the skewness parameter $\alpha $, is found for both
materials --- occurring near $H\!=\!2.5$~T and $4$~T for $\kappa $-CuNCS and 
$\kappa $-Br, respectively. These values are of the order of the respective
irreversibility fields extracted from low-temperature magnetization
measurements,\cite{Sasaki98,Hagel97} and much larger than the 3D-2D
crossover fields for electromagnetically coupled layers of pancake vortices
that have been reported for these materials (e.g. $H_{\mathrm{2D}}\! \sim
\!10^{-2}$~T for $\kappa $-CuNCS\cite{Lee-ET-PRL97}). Furthermore, the
observation of positive values of $\alpha $ in $\kappa $-Br for magnetic
fields up to $H\!=\!2$~T at the higher temperature of $T\!=\!1$~K, reflects
a robustness of the 3D vortex lattice to thermal fluctuations of pancake
vortices in the individual SC layers. Analytical calculations based on a 2D
Ginzburg-Landau theory,\cite{ZM99,Maniv01} which agree well with numerical
Monte Carlo simulations,\cite{HuMacDonald93-94} show, for material
parameters characteristic of $\kappa $-CuNCS [see Fig.~3(b) in Ref.~%
\onlinecite{Maniv2004}], that melting of an ordered vortex lattice at $%
T\!=\!0.1T_{c}\! \approx \!1$~K, in each individual 2D SC layer, occurs
around $H\!=\!0.4H_{c2}\! \approx \!2$~T. Moreover, the irreversibility
field,\cite{Sasaki98} which may be associated with a critical field for
depinning of vortices, follows closely the 2D vortex-lattice melting field
[also shown in Fig.~3(b) of Ref.~\onlinecite{Maniv2004}]. Numerical
simulations of the skewness parameter $\alpha $, performed for weakly
coupled layers of pancake vortices,\cite{SchneiderPRB95} show that a 3D-2D
crossover reduces $\alpha $\ to about 30\% of its (positive) low-field
value, whereas melting of the constituent 2D vortex lattices causes the sign
of $\alpha $ to abruptly change from positive to negative. Since the values
of $\alpha $ shown in Fig.~3 below the order-disorder transition fields are
larger than that predicted for a disordered multilayer of 2D vortex lattices
(i.e. $\alpha \! \simeq \!0.3$),\cite{SchneiderPRB95} it is conceivable that
no 3D-2D crossover takes place in the low-field region. Instead, it appears
that the only order-disorder transition that occurs in the entire field
range investigated, is melting of a partially-ordered 3D vortex lattice.
This conclusion is consistent with the single first-order vortex lattice
melting transition predicted in a Monte-Carlo study of the Lawrence-Doniach
model for a layered superconductor.\cite{HuMaC97}

The apparent absence of a low-field 3D-2D crossover in our measurements
suggests that the observed 3D partially-ordered vortex lattice is probably
stabilized by strong pinning due to a small concentration of defects. This
scenario is supported by the proximity of the irreversibility (depinning)
transition to the melting transition, previously observed for $\kappa $%
-CuNCS.\cite{Maniv2004} However, it should be noted that the sharp sign
change of $\alpha $ observed for both materials, could be associated with a
transition to a vortex-glass phase,\cite{Menon06,Maniv2011} rather than a
vortex-lattice melting transition. \ 

The primary source of flux-line pinning may be a consequence of the high
rate at which the samples{\ }were pre-cooled when loaded into the dilution
refrigerator. During the pre-cooling stage, the sample is cooled from room
temperature to $4.2$~K in about 15 to 20 minutes, passing through the
glass-like structural transition at $T_{g}\sim 75$ K, which -- for the $%
\kappa $-phase ET salts with polymeric anions Cu(NCS)$_{2}$ and Cu[N(CN)$%
_{2} $]Br -- is related to a certain degree of disorder in the orientational
degrees of freedom of the ET molecules' ethylene endgroups (EEG) (see Ref.~%
\onlinecite{Toyota07} for an overview). The EEG orientations are thermally
disordered at room temperature, and for kinetic reasons, become frozen in a
non-equilibrium configuration depending on the cooling rate at $T_{g}$.\cite%
{Muller02} The pre-cooling rate in our experiments was about $20$ K/min,
which corresponds to a random lattice potential associated with structural
disorder in $3$ to $4$ $\%$ of the sample. \cite{Hertmann14} Clustering of
metastable EEG configurations in this volume fraction of the sample and into
domain sizes exceeding the in-plane coherence length,\cite{Yoneyama04} is a
likely source of the strong flux-line pinning that accommodates the
partially-ordered vortex lattice. \bigskip

This research was supported by TECHNION/V.P.R FUND, and by EuroMagNET under
the EU contract No.\ 228043, and by the Deutsche Forschungsgemeinschaft
(DFG) within the collaborative research center SFB/TR49. J. E. Sonier
acknowledges support from the Natural Sciences and Engineering Research
Council of Canada. T.M. is grateful to T. Sasaki for stimulating discussions
regarding relevant experimental aspects of the paper, and to Dingping Li and
B. Rosenstein for invaluable discussions of theoretical aspects.

\end{document}